\documentstyle[preprint,aps]{revtex}
\begin{document}
\def\sla#1{\rlap\slash #1}
\draft
\title{Rho meson form-factos in the 
null-plane phenomenology 
\footnote{International Workshop on Hadrons Physics 96, 
Topics on the Structure and Interaction of 
Hadronic Systems, Brazil, Editors: Erasmo Ferreira et all.,
World Scientific 1997.}}
\author{J.P.B.C. de Melo$^a$ and
T. Frederico$^b$}
\address{
$^a$ Instituto de F\'\i sica, Universidade de S\~ao Paulo, \\ 01498-970
S\~ao Paulo, S\~ao Paulo, Brazil. \\
$^b$ Departamento de F\'\i sica, ITA, 
Centro T\'ecnico Aeroespacial, \\
12.228-900 S\~ao Jos\'e dos Campos, S\~ao Paulo, Brazil.} 
\date{\today} 
\maketitle

\begin{abstract} 
The constituent quark rho meson electromagnetic form-factors 
are calculated, with covariant and null-plane 
approaches with the same model. 
The null-plane formalism produces the breakdown of the rotational 
symmetry for the one-body current operador, wich is 
investigated by comparing the numerical results in both 
approaches. This allows to choose the appropriate null-plane 
prescription, among the several ones, to evaluate the rho meson 
form-factors. \\ 
\end{abstract} 
\narrowtext

Since Dirac \cite{dirac}, it is known that the light-front hypersurface
given by $x^+=x^0+x^3=0$ (null-plane) is suitable for
defining the initial state of a relativistic system. 
Relativistic models with null-plane wave-functions have becoming 
widely used in particle phenomenology \cite{teren}. 
It is well known that \cite{Frankfurt81,Grach84} 
,the $J^{+}(=J^{0}+J^{3})$ componente of the 
electromagnetic current, looses its rotational invariance 
for a spin1 composite system.  
The matrix elements are computed with a null-plane wave-function 
\cite{Chung88,Frankfurt93} in the Breit-Frame, where the vector 
component of the momentum transfer is in the x-direction 
$(q^{\mu}=(0,q_{x},0))$. If rotational symmetry around x-axis 
is valid, we should have $J^{+}_{zz}$$=$$J^{+}_{zz}$, 
where subscripts are the polarization, taken in the cartesian 
instant-form spin basis \cite{Frankfurt93}. 
Such requirement is called angular condition 
\cite{Frankfurt81,Grach84}. 
The breakdown of rotational symmetry, implies 
that, it does not exist an unique way to extract electromagnetic 
form factors from the matrix elements of $J^{+}$, for 
composite systems with spin1. 
In the literature, the are several extraction schemes for 
the form-factors, by using different linear combinations of 
$J^{+}$ matrix 
elements \cite{Grach84,Chung88,Frankfurt93,Brodsky92}. 

The covariant approach does not violate rotational 
symmetry, and we compare it with the results obtained for 
the $\rho$-meson form-factors with different extraction schemes 
in the null-plane. The $\rho$-meson 
charge $(G_0$$(q^{2}))$, magnetic $(G_1$$(q^{2}))$ 
and quadrupole $(G_2$$(q^{2}))$ form-factors \cite{Chung88}, 
are calculated in the impulse approximation. 
The amplitude for the photon absorption is given by the 
Feynman triangle-diagram, with the photon leg attached to one 
of the quarks. 
The vertex for the $\rho$-$q\bar{q}$ coupling is given by 
\begin{equation}
\Gamma^\mu (k,k') = \gamma^\mu -\frac{m_\rho}{2}
 \frac{k^\mu+k'^\mu}{ p.k + m_\rho m -\imath \epsilon}  \ .
\label{eq:rhov}
\end{equation}
Where, the $\rho$-meson  is on-mass-shell, with mass $m_{\rho}$, 
and its, four is $p^{\mu}$=$k^{\mu}$ - $k^{'\mu}$, 
the quark momenta are given by $k^{\mu}$ and $k^{'\mu}$. 
The constituinte quark mass is m. 
For a on-mass-shell quark, Eq.1 reduces to a S-wave vertex 
\cite{Frankfurt81}.
The impulse approximation to $J^{+}$, is given by the Feynman 
triangle-diagram, where the constituinte quark is a pointlike 
particle, 
\begin{eqnarray}
J^+_{ji}&=&\imath  \int\frac{d^4k}{(2\pi)^4}
 \frac{Tr[\epsilon^{'\alpha}_j \Gamma_{\alpha}(k,k-p_f)
(\sla{k}-\sla{p_f} +m) \gamma^{+} 
(\sla{k}-\sla{p_i}+m) \epsilon^\beta_i \Gamma_{\beta}(k,k-p_i)
(\sla{k}+m)]}
{((k-p_i)^2 - m^2+\imath\epsilon) 
(k^2 - m^2+\imath \epsilon)
((k-p_f)^2 - m^2+\imath \epsilon)}
\nonumber \\ & &\times \Lambda(k,p_f)\Lambda(k,p_i) \ ,
\label{eq:tria}
\end{eqnarray}
where the polarization four-vector are $\epsilon^{'\alpha}_{i}$ and 
$\epsilon^{'\beta}{_j}$ for the initial and final state, 
respectively. 
The $\rho$-meson four momenta are given by $p_{i(f)}$, 
$i(f)$ stands for initial (final). The calculation is performed in the 
Breit-frame. 
The regularization function, 
$ \Lambda(k,p)=N/((k-p)^2-m_{R}+  i \epsilon)^2 $, was chosen 
to turn Eq.2 finite. Such regulator gives a null-plane wave-function 
similar to the one recently proposed for the pion 
\cite{Shakin95}, $N$ is obtained by $G_{0}=1$.

The null-plane calculation, corresponds to integrate over $k^{-}$ 
\cite{Frederico92}. The pair diagrams are not present and the pole 
in $k^{-}$ complex-plane wich contributes is 
$k^{-}=(k_{\perp}^2+m^2- \imath \epsilon)/k^{+}$, with 
$p^{+} > k^{+} > 0 $. The energy of the $\rho$-meson in 
the Breit-frame is $p^{+}=p^{0}$. Such pole corresponds to a spectator 
quark in the process of photon absortion being on-mass-shell. 
The null-plane wave-function  of the $\rho$-meson appears after 
the substituion of the on-mass-shell momentum of the spectator quark 
in the quark-propagador and in the regulator. The resulting expression 
in the center of mass system gives,
\begin{eqnarray}
\Phi_i(x,\vec k_\perp)=\frac{N^2}{(1-x)^2(m^2_\rho-M_0^2)
(m^2_\rho- M^2_R)^2} 
\vec \epsilon_i . [\vec \gamma -  \frac{\vec k}{\frac{M_0}{2}+ m}] \ ,
\label{eq:npwf}
\end{eqnarray}
where, $x=k^{+}/p^{+}$. 
The free square mass of the virtual quark-antiquark system is given by 
$M_{0}^{2}=(k_{\perp}^2+m^2)/x+((\vec{p}-\vec{k}))_{\perp}^2+m^2)/(1-x)-
p_{\perp}^2$. 
The function $M_{R}^2$ is given by 
$M_{R}^{2}=(k_{\perp}^2+m^2)/x+((\vec{p}-\vec{k}))_{\perp}^2+m_{R}^2)/(1-x)-
p_{\perp}^2$. We left out the phase-space factor $1/(1-x)$.

The $\rho$-meson mass is $0.77$ GeV and the composite 
wave-function corresponds to a bound state. The 
parameters $m=0.43$ GeV and $m_{R}= 1.8$ GeV, are adjusted 
with the covariant calculation such that the mean square 
radius is about $0.37$ $fm^2$ and $G_{2}(q^2 \sim 5$ GeV) $\sim -0.25$.

Those values were obtained with point-like 
constituintes quarks in a model with 
one-gluon exchange at short distances and linear confinement 
at large distances \cite{Cardarelli95}. 

We compare the results of the different prescriptions in the 
null-plane with the covariant ones. 
In Fig.1, we observe that the charge form-factor, 
$G_{0}$, is sensitive to the different prescriptions 
in the null-plane mainly above 3 $GeV^2$. 
The (GK) \cite{Grach84} prescription gives results 
in agreement with the covariant calculation, while (BH) 
results are about 30 \%  below at higher $q^2$. 
We present the results for the magnetic moment form-factor in Fig.2.

The differences with the covariante  calculation are not so 
pronounced and at small momentum transfer (FFS) cite{Frankfurt93} 
prescription has a value about 15 \% higher than the covariant result. 
The relativistic effects in the model gives $G_{2}$, and 
thus it is more sensitive to different prescriptions. 
In Fig.3, the $G_{2}$ calculated in the null-plane 
with prescriptions given by (CCKP) \cite{Chung88} 
and (BH) \cite{Brodsky92} are 20 \% lower than the covariant 
result. We conclude that, in the scale of the $\rho$-meson 
bound state, tuned by a parametrization wich reproduces the size and 
the quadrupole form-factor, of an effective constituent quark model, 
wich bodies gluon exchange and confinement; the prescription for 
a null-plane calculation as given by the work of Grach and 
Kontratyk \cite{Grach84} shows the better agreement with a 
covariant calculation.


\begin{figure}[h]
\vspace{10.0cm}
\includegraphics{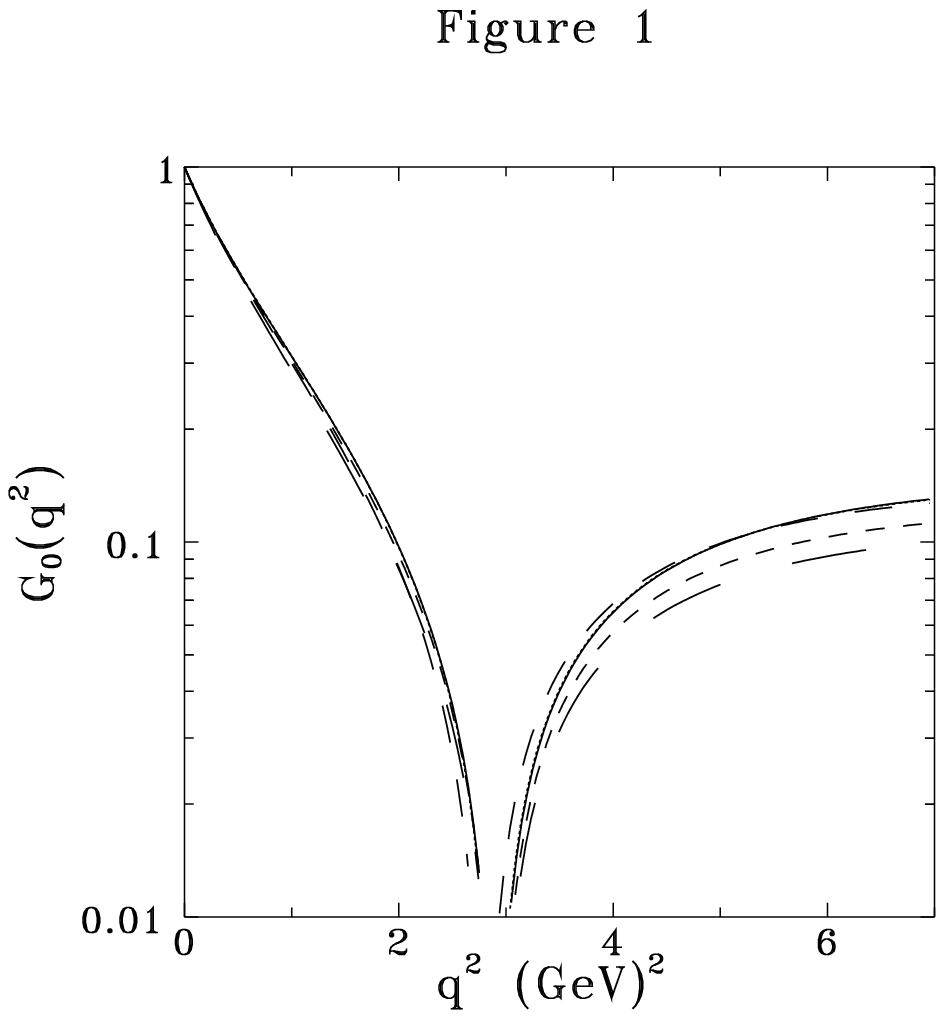}
\caption{ Charge form-factor $ G_0(q^2) $ 
for the $\rho$-meson as a function of $q^2$, calculated
with covariant and light-front schemes.
The solid line is the covariant calculation.
Results for the different light-front
extraction schemes, (GK) (dotted line)
(it is not possible to distinguish from the covariant calculation), 
(CCKP) (short-dashed), Ref. [8]
(FFS) (dashed) and (BH) (long-dashed).}
\label{fig1}
\end{figure}

\newpage
.

\begin{figure}[h]
\vspace{10.0cm}
\includegraphics{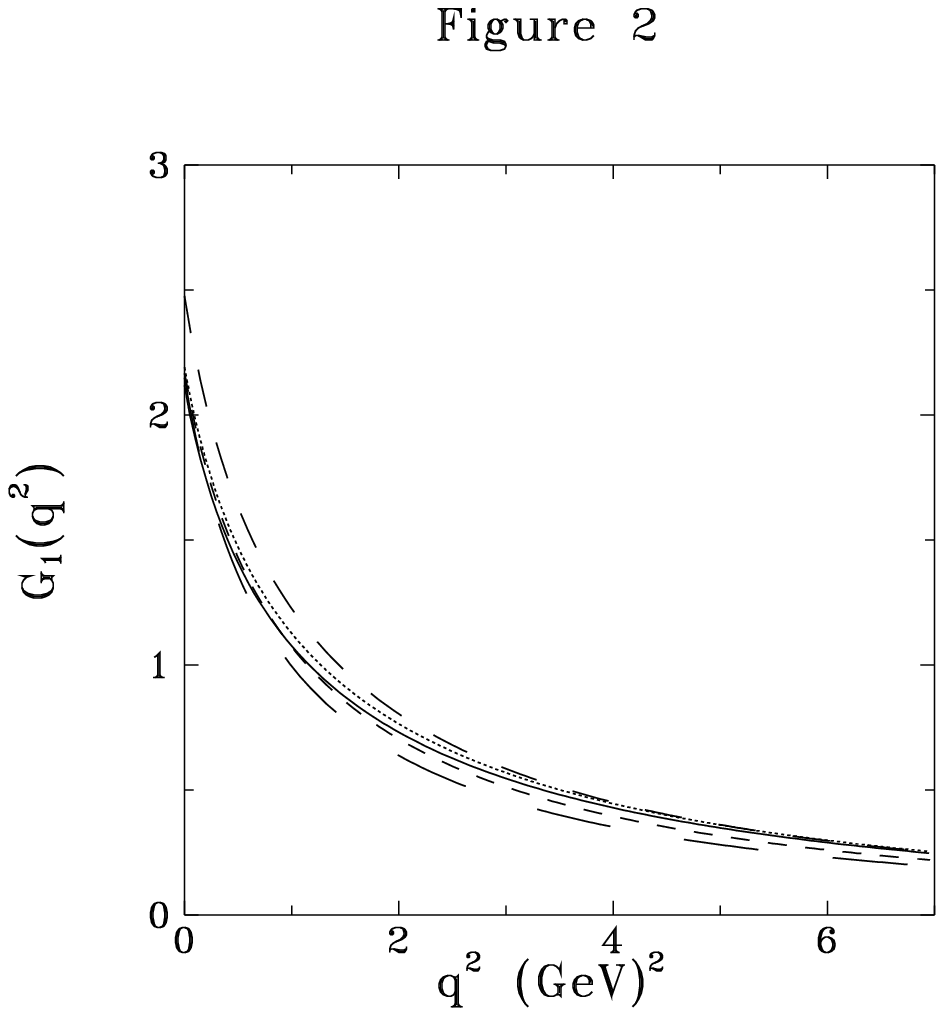}
\caption{ Magnetic form-factor $ G_1(q^2) $ 
for the $\rho$-meson as a function of $q^2$, calculated
with covariant and light-front schemes.
The curves are labeled according to Fig.1.}
\label{fig2}
\end{figure}

\newpage
.

\begin{figure}[h]    
\vspace{10.0cm}
\includegraphics{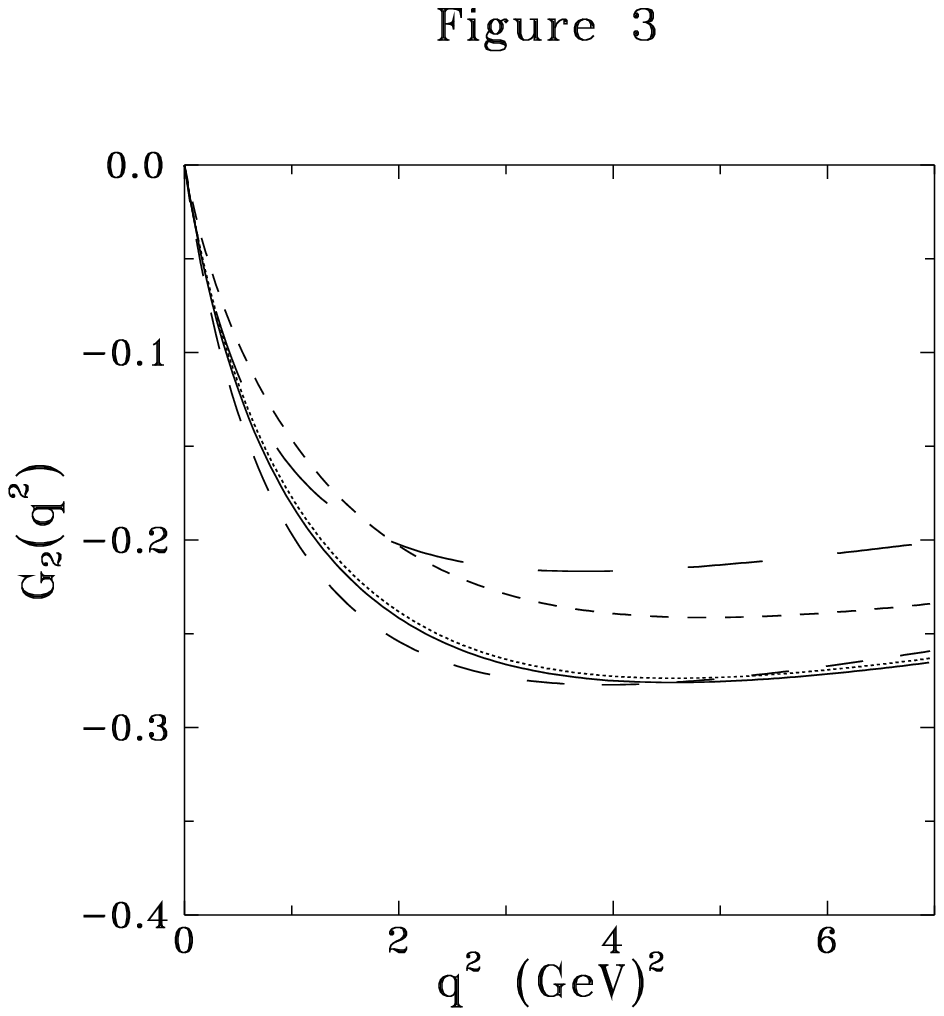}
\caption{ Magnetic form-factor $ G_2(q^2) $
for the $\rho$-meson as a function of $q^2$, calculated
with covariant and light-front schemes.
The curves are labeled according to Fig.1.}
\label{fig3}
\end{figure}

\end{document}